\documentclass[journal]{IEEEtran}
\usepackage{balance}

\usepackage{cite}

\ifCLASSINFOpdf
   \usepackage[pdftex]{graphicx}
\else
   \usepackage[dvips]{graphicx}
\fi
\usepackage{amsmath}
\usepackage{amsfonts}

\DeclareMathOperator*{\argmin}{arg\,min}
\usepackage{algorithmic}
\ifCLASSOPTIONcompsoc
 \usepackage[caption=false,font=normalsize,labelfont=sf,textfont=sf]{subfig}
\else
 \usepackage[caption=false,font=footnotesize]{subfig}
\fi
\begin{document}
%
\title{Spectral Graph Clustering for Intentional Islanding Operations in Resilient Hybrid Energy Systems}
%
%
%

\author{Jiaxin~Wu,~\IEEEmembership{Student Member,~IEEE}, Xin~Chen,~\IEEEmembership{Member,~IEEE}, Sobhan~Badakhshan~\IEEEmembership{Student Member,~IEEE}, Jie~Zhang,~\IEEEmembership{Senior Member,~IEEE,}
        and~Pingfeng~Wang,~\IEEEmembership{Member,~IEEE}
\thanks{J. Wu, X. Chen and P. Wang are with the Department
of Industrial and Enterprise Systems Engineering, University of Illinois Urbana-Champaign, Urbana,
IL 61801, USA, e-mail: pingfeng@illinois.edu.}
\thanks{S. Badakhshan, J. Zhang are with the Department of Mechanical Engineering and (Affiliated) the Department of Electrical and Computer Engineering, University of Texas at Dallas, Richardson,
TX 75080, USA.}
}

%
%

\markboth{}%
{Shell \MakeLowercase{\textit{et al.}}: Bare Demo of IEEEtran.cls for IEEE Journals}
%



\maketitle

\begin{abstract}
Establishing cleaner energy generation therefore improving the sustainability of the power system is a crucial task in this century, and one of the key strategies being pursued is to shift the dependence on fossil fuel to renewable technologies such as wind, solar, and nuclear. However, with the increasing number of heterogeneous components included, the complexity of the hybrid energy system becomes more significant. And the complex system imposes a more stringent requirement of the contingency plan to enhance the overall system resilience. Among different strategies to ensure a reliable system, intentional islanding is commonly applied in practical applications for power systems and attracts abundant interest in the literature. In this study, we propose a hierarchical spectral clustering-based intentional islanding strategy at the transmission level with renewable generations. To incorporate the renewable generation that relies on the inverter technology, the frequency measurements are considered to represent the transient response. And it has been further used as embedded information in the clustering algorithm along with other important electrical information from the system to enrich the modeling capability of the proposed framework. To demonstrate the effectiveness of the islanding strategy, the modified IEEE-9 bus and IEEE-118 bus systems coupled with wind farms are considered as the test cases.
\end{abstract}

\begin{IEEEkeywords}
intentional islanding, disruption management, resilience, renewable.
\end{IEEEkeywords}

%
\IEEEpeerreviewmaketitle

\section{Introduction}
%
%
%
%
\begin{figure*}
    \centering
    \includegraphics[width = 0.8\linewidth]{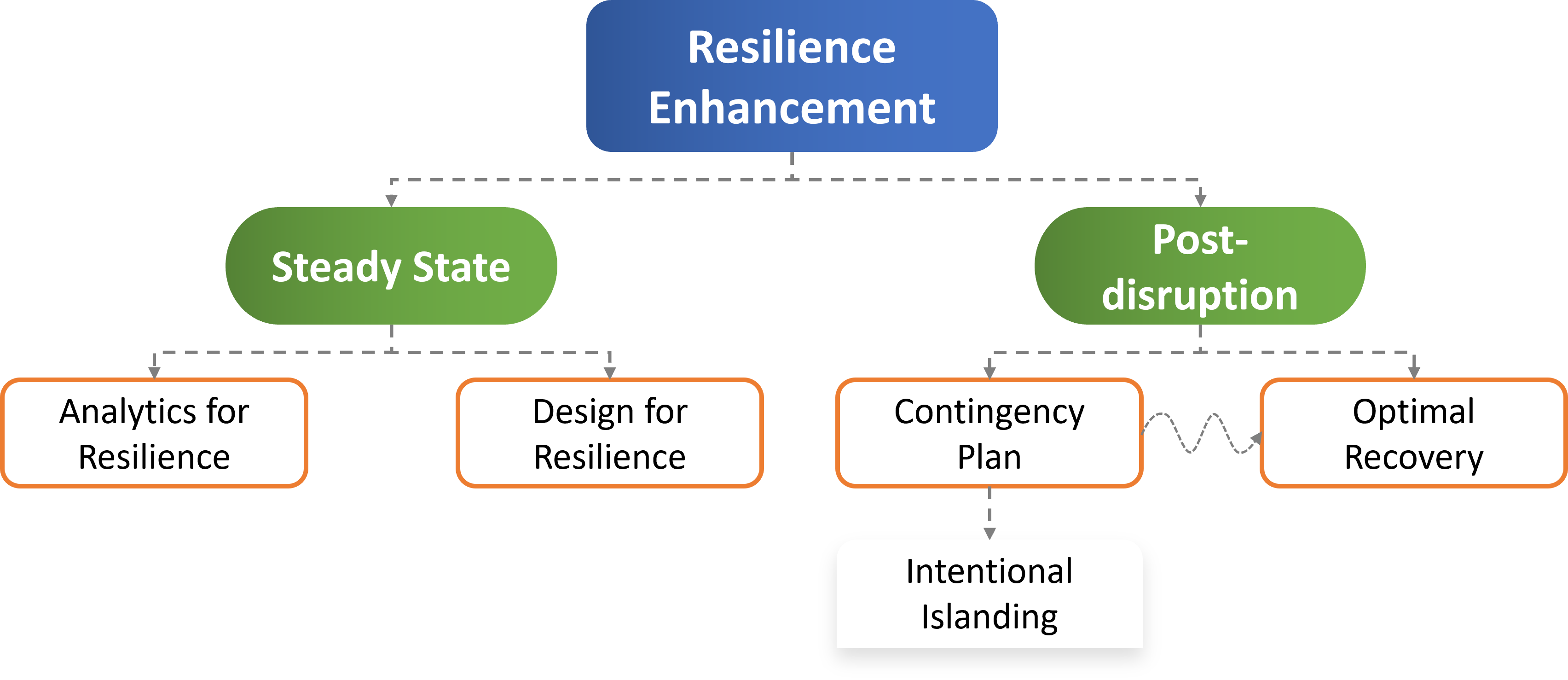}
    \caption{The studies for resilience enhancement can be categorized into two groups, while the intentional islanding is one of the contingency plans after disruptions.}
    \label{fig:tree_plot_introduction}
\end{figure*}
\IEEEPARstart{W}{ith} the increases on both scale and complexity, hybrid energy systems (HESs) consisting of various types of generation technologies, become more vulnerable towards disruptive events. Such vulnerability therefore drives the research efforts that could lead to robust and resilient HESs. Questions arise, for instance, how to efficiently design a large-scale system that can resist potential external disruptions, or how can the decision-maker evaluate the uncertain dynamic behavior of the HES undergoing different disruptive events? Also, for large-scale HESs, mitigating the cascading effects after occurrences of disruptive events is a crucial task for attaining reliable system operations. To quantify the system's performance during disruption, or to comprehend the system's capability toward uncertain disruptive scenarios, researchers have adopted the term "resilience" from the ecology field \cite{DeAngelis1980}. Different from the terminology of system reliability, in which the time-dependent degraded system performance and the possibility of failure are studied, the resilience metric is utilized to complement the analysis of real-time system behavior. Based on the U.S. Department of Defense report, a resilient system should not only withstand impacts of disruptive events but also need to acquire the capability of self-healing from damages \cite{Goerger2014}. Thus, to realize a resilient HES through design or operational management strategies, the stakeholders need to tackle challenges in three folds: (1) with the possibility of external or internal disruptions, how should the system proactively detect the occurrence of abnormalities; (2) how large is the bandwidth of the HES for withstanding adversarial impacts; (3) how quick the HES can self-recover to its nominal state \cite{Walker2004}. 

Motivated by the challenges in the previously mentioned three aspects, different frameworks have been proposed to help the HES establish self-healing capability after system disruptions therefore achieve an adequate resilience level. Here, we categorize the research efforts about engineering resilience in terms of the temporal stages of the proposed frameworks, i.e. before and after the disruptions as shown in Fig.~\ref{fig:tree_plot_introduction}. During the steady-state when the system operates under its nominal performance, it's important to obtain a universal analytic framework to assess the resilience level of ICIs, by considering the potential disruptive events \cite{Sharma2018}. Moreover, researchers have proposed to use a probabilistic graphical model such as Bayesian Network to gauge the system resilience level dynamically with evolving system behavior \cite{Yodo2016}. It is also beneficial to optimize the system resilience during the initial design stage, where various resilience enhancement methods can be considered: setting up a preventive or corrective maintenance schedule, establishing appropriate redundancies, or conducting proactive system health management \cite{Compare2014}. Although comprehensive pre-disruption frameworks have been proposed to either quantify the resilience of HESs or introduce methodologies to improve the system resilience in advance, how a system should behave after disruptions thus to preserve resilient performance is still unknown. And without an appropriate post-disruption guiding/response framework, cascading failures induced by severe disruptions are inevitable, especially in closely coupled network systems. As a result, it's required to study suitable recovery strategies to ensure system resilience after disruptions.

To tackle the challenges of achieving the self-healing capability of a resilient system, several real-time operational frameworks have been proposed to guide how the system should behave after disruptive events. For instance, researchers try to attain a resilient operational framework by incorporating optimal scheduling for repair tasks \cite{Ouyang2017, WU2021107836}, as well as repair resources \cite{wu2021post} with uncertainties, forming self-sustainable subsystems \cite{Chen2016}, and guided recovery through control strategies \cite{Wu2019}. All aforementioned studies focus on solving for the optimal decisions of how to utilize the existing resources or back-ups to recover the HES thus mitigating the effects of disruptions, on the basis of having a contingency plan such as the network reconfiguration beforehand. In other words, during the post disruption stage, the self-recovery capability is realized in two steps: an appropriate emergency response e.g. system reconfiguration, followed by performing optimal restorations. As a result of the necessary sequential order, setting up a reliable system reconfiguration plan before conducting optimal recovery is vital for ensuring system resilience. And in the power system field, the system reconfiguration is widely used as a contingency plan i.e. the intentional islanding strategy for a system undergoing disturbances to mitigate cascading effects \cite{25627, 6742636, 7878584, 8472795} or to form robust microgrid operations \cite{7553473, 7802603, 7271061, 8839824}.

The main objective of conducting intentional islanding is to derive the optimal plan for allocating buses of the HES into appropriate disconnected subsystems. Thus the system can isolate failures and sustain nominal performance with the presence of disruptions. Such a contingency plan coincides with the goal of clustering algorithms. Therefore, proposed intentional islanding strategies usually use the graph clustering method as the building block. To briefly review existing frameworks, for example, Esmaeilian et al.~\cite{Esmaeilian2017} have proposed a spectral clustering based intentional islanding strategy to regulate the systems after disruptions, considering solely the system power flow as the major performance criterion. Moreover, to mitigate the effect of the presumption on the number of islands after disruptions, Sanchez-Garcia et al.~\cite{Sanchez-Garcia2014} have utilized the unsupervised hierarchical clustering technique to conduct the system partitioning. Dabbaghjamanesh et al.~\cite{Dabbaghjamanesh2020} have improved the robustness of the intentional islanding strategy by considering multiple electrical information embedded in the system along with the dynamic power flow measurement in the algorithm. However, all aforementioned works have focused on power systems only with synchronous generators and usually rely on a specific set of system measurements. As for modern power girds with a high level of renewable penetrations, the system performance will be largely influenced by the behaviour of the inverter-based generation. Hence, how to perform the intentional islanding operation for power systems with renewable penetration is still a crucial challenge for improving the overall system resilience. Besides, nowadays HESs usually undergo expansion plannings, and with the additional heterogeneous components, the online measurement information can vary along the way \cite{SADEGHI20171369}. Thus, the stake holder needs a universal intentional islanding framework that can unify the varying information to make robust decisions.

In this paper, we demonstrate a new intentional islanding method based on the clustering algorithm for HES with various levels of renewable penetrations to enable the self-recovery capability. Our proposed framework utilizes the hierarchical spectral clustering technique based on both systems' static and dynamic information. And the advantages can be concluded in two folds: (1) the algorithm can work not only for synchronous generators but also for HESs with inverter based renewable generations; (2) the intentional islanding framework is universal that it is applicable for different HESs with various electrical information. Also we consider the modified IEEE-9 bus and IEEE-118 bus test systems as case studies to illustrate the effectiveness of the proposed method.

The remaining parts of the paper is organized as follows: Sec.~\ref{sec:methodologies} introduces the proposed intentional islanding methodology in detail, where the spectral graph clustering algorithm is coupled with the Grassmann manifold technique to accommodate the heterogeneous information extracted from the system. Sec.~\ref{sec:case_study} presents case studies results based on the IEEE-9 bus and IEEE-118 bus test systems to validate the proposed islanding framework. And Sec.~\ref{sec:conclusion} summarizes the study along with discussion about future research directions.

\section{Clustering for Intentional Islanding}\label{sec:methodologies}
The intentional islanding is usually realized by conducting node clustering for the power system represented as a graph. As for the clustering methods, the K-Means algorithm is widely used for identifying inherent patterns of high dimensional data. The algorithm assumes that each sample point belongs exclusively to one latent cluster. And it assigns the data point $X_j$ to the cluster $S_i$ in terms of minimizing the overall Euclidean distance from the cluster center $\mu_i$ to each point. And the overall objective for a dataset with $n$ samples can be expressed as:
\begin{equation}
    \argmin_S \sum_{j=1}^n\sum_{X_j \in S_i}\vert\vert X_j-\mu_i\vert\vert ^2.
\end{equation}
Although the K-Means algorithm has adequate clustering performance for tabular data, it cannot be directly applied to graph-structured data, e.g., HESs and other types of cyber-physical systems. Since the Euclidean distance cannot directly reflect the difference between the nodes in the network system. To remedy this limitation of the K-Means algorithm, spectral graph theory can be incorporated. To be self-contained, we introduce the spectral clustering algorithm as well as the mechanism for embedding the graphical information of the intentional islanding strategy in the following sections.

\subsection{Spectral Graph Theory}\label{sec:spectral_theory}
Let a graph $\mathcal{G}=(\mathcal{V},\mathcal{E})$ with vertex set $\mathcal{V}$ and edge set $\mathcal{E}$ represent the interconnected system. To be specific, for a $N$-bus power system, the corresponding graphical notation can be written as $\mathcal{V}:=\{1,2,\dots,N\}$ and $\mathcal{E} := V\times V$. In the graph, the buses connecting to loads and generations are the vertices while the transmission lines are the edges. Moreover, based on the connection status between each pair of buses, a square connectivity matrix $W^t_{ij}\in \mathbb{R}^{V\times V}$ can be constructed as:
\begin{align}
    W^t_{ij} = 
    \begin{cases}
    1 &\text{if}\; (i,j)\in{\mathcal{E}}, \\
    0 &\text{otherwise}.
    \end{cases}
\end{align}
Notice that the $W^t_{ij}$ is symmetric and follows the convention that all diagonal entries equal to zero, i.e., no self-loop. The connectivity matrix solely indicates the topological information of the graph-structured system. Besides the connectivity matrix, the power system requires other matrices to embed the electrical information it preserves, such as the power flow and resistance/reactance. Thus, by using $W^t_{ij}$ as an analogy, matrices for the electrical properties are defined as:
\begin{align}
    W^p_{ij} = 
    \begin{cases}
    \frac{\vert P_{ij} \vert+ \vert P_{ji}\vert}{2} &\text{if}\; (i,j)\in{\mathcal{E}}, \\
    0 &\text{otherwise},
    \end{cases}
\end{align}
where $P_{ij}$ and $P_{ji}$ are the active power flow on each transmission line considering the line loss. And the matrix for admittance of the system can be obtained as:
\begin{align}
    W^a_{ij} = 
    \begin{cases}
    \frac{1}{\sqrt{R_{ij}^2 + X_{ij}^2}} &\text{if}\; (i,j)\in{\mathcal{E}}, \\
    0 &\text{otherwise},
    \end{cases}
\end{align}
where $R_{ij}$ and $X_{ij}$ are the resistance and reactance of the transmission lines, respectively. Similar to the connectivity matrix $W^t_{ij}$, the power flow matrix $W^p_{ij}$ and admittance matrix $W^a_{ij}$ are also symmetric with all diagonal entries being zeros.

The motivation of establishing the aforementioned three weight matrices is that they measure how strong a line connection would be when conducting intentional islanding by clustering nodes. In other words, disconnecting a transmission line with a large weight to form islands after clustering will lead to a large penalty in the objective of the clustering algorithm. And this behavior is desirable for intentional islanding to prevent cascading failures. Since buses directly connected are more likely to stay in the same cluster with similar operating conditions. Moreover, a larger admittance means a shorter electrical distance, and the buses connected by a transmission line with a small electrical distance should be in the same cluster after the islanding. Similarly, transmission lines carrying large power flow should not be cut and form isolated islands, which could lead to a large imbalance in load/generation and load shedding. Notice that the electrical information used in the intentional islanding algorithm are not limited to the aforementioned three matrices. Based on the prior knowledge of the decision maker, other important online measurements from the HES can also be used to represent the system status and thus treated as the matrical inputs of the algorithm.

In order to cluster nodes based on introduced weight matrices and fulfill the intentional islanding task, we need to further derive the Laplacian matrix of the graph. The Laplacian matrix embeds the nodal information from the graph system into a pseudo-tabular form. Then clustering algorithms, which are developed based on the structured data, can use the derived Laplacian matrix as the input. And it is defined as $L = D - W$ where $D$ is the degree matrix of the graph, which is a diagonal matrix obtained from $D=diag(\sum_{j}W_{ij})$. That is, the degree matrix $D$ has all zeros on its non-diagonal entries, while the diagonal elements are the row sum of the weight matrix. In this way, each element of the Laplacian matrix has the form
\begin{align}
    L_{ij} = 
    \begin{cases}
    d_i &\text{if}\;i=j,\\
    -w_{ij} &\text{if}\;(i,j)\in{\mathcal{E}},\\
    0 &\text{otherwise},
    \end{cases}
\end{align}
where $d_i$ is the degree of node $i$. Usually, we will work on the normalized Laplacian matrix, which is scale independent and has better results for clustering. The normalized Laplacian matrix is calculated as $L_n = D^{1/2}LD^{-1/2}$. After normalization, the entries of the Laplacian matrix are
\begin{align}
    [L_n]_{ij} = 
    \begin{cases}
    1 &\text{if}\;i=j,\\
    \frac{-w_{ij}}{\sqrt{d_id_j}} &\text{if}\;(i,j)\in{\mathcal{E}},\\
    0 &\text{otherwise}.
    \end{cases}
\end{align}
One characteristic of the Laplacian matrix is that it has the spectral information of the graph. We can perform eigendecomposition on the Laplacian matrix and treat the eigenvectors as the spectral embeddings. Then by conducting clustering on these obtained spectral embeddings, we can discover the latent clusters for the network system. The overall roadmap of such a clustering technique is to use the first $K$ eigenvectors of the Laplacian matrix as the geometric coordinates of the $N$ data points defined in $\mathbb{R}^K$. Then these coordinates can be treated as in Euclidean space and be clustered by standard clustering techniques such as the K-Means algorithm. Notice that all spectra of the normalized graph Laplacian, i.e., its eigenvalues, always lie between zero and two. This magnitude is desirable when calculating the nodal similarities based on the corresponding eigenvectors. 

\subsection{Hierarchical Spectral Clustering}
In Sec.~\ref{sec:spectral_theory}, we have briefly introduced the spectral clustering algorithm for graph systems based on the K-Means algorithm. However, one drawback of the K-Means based spectral clustering is that the number of clusters, $K$, needs to be predefined. As for intentional islanding to avoid cascading failures, the number of islands cannot be easily assumed beforehand but need to be determined based on the online operational condition. To overcome this limitation of the K-Means algorithm, this study utilizes hierarchical clustering on the spectral domain of the graph.

Different from the K-Means algorithm, which directly outputs the partitioning results based on the given number of clusters $K$ and omits the inner connection between the nodes in the same cluster, the hierarchical clustering provides partitioning results with finer intra-cluster detail. The results of the hierarchical clustering are usually interpreted by the tree-like dendrogram, where each leaf is the individual node in the graph. The leaves of the dendrogram can also be viewed as clusters with a cardinality of one. And the goal of the clustering algorithm is to merge the closest clusters into a bigger cluster at each step following a bottom-up approach. To determine the best merging decision, the Wade linkage criterion~\cite{de2015feature} is used as the metric in this study:
\begin{equation}
    d(u,v)=\sqrt{\frac{|v|+|s|}{T}d(v,s)^2+\frac{|v|+|t|}{T}d(v,t)^2-\frac{|v|}{T}d(s,t)^2}
\end{equation}
where $u$ is a merged cluster from subclusters $s$ and $t$, $v$ is a unused cluster in parallel with $s$ and $t$, $|*|$ is the cardinality operator, and $T$ equals the sum of $|s|$, $|t|$, and $|v|$. The Wade linkage criterion is a variance minimization metric and picks two clusters to form one that leads to the lowest sum of squared differences within the clusters. This variance minimizing characteristic makes the Wade linkage criterion similar to the objective of the traditional K-Means algorithm, but in a hierarchical setting. Again, with the Wade linkage criterion, the goal is to merge two clusters with the smallest $d(u,v)$ until all nodes are merged into a single cluster, which is the root of the dendrogram.

\begin{figure}
    \centering
    \includegraphics[width = 0.8\linewidth]{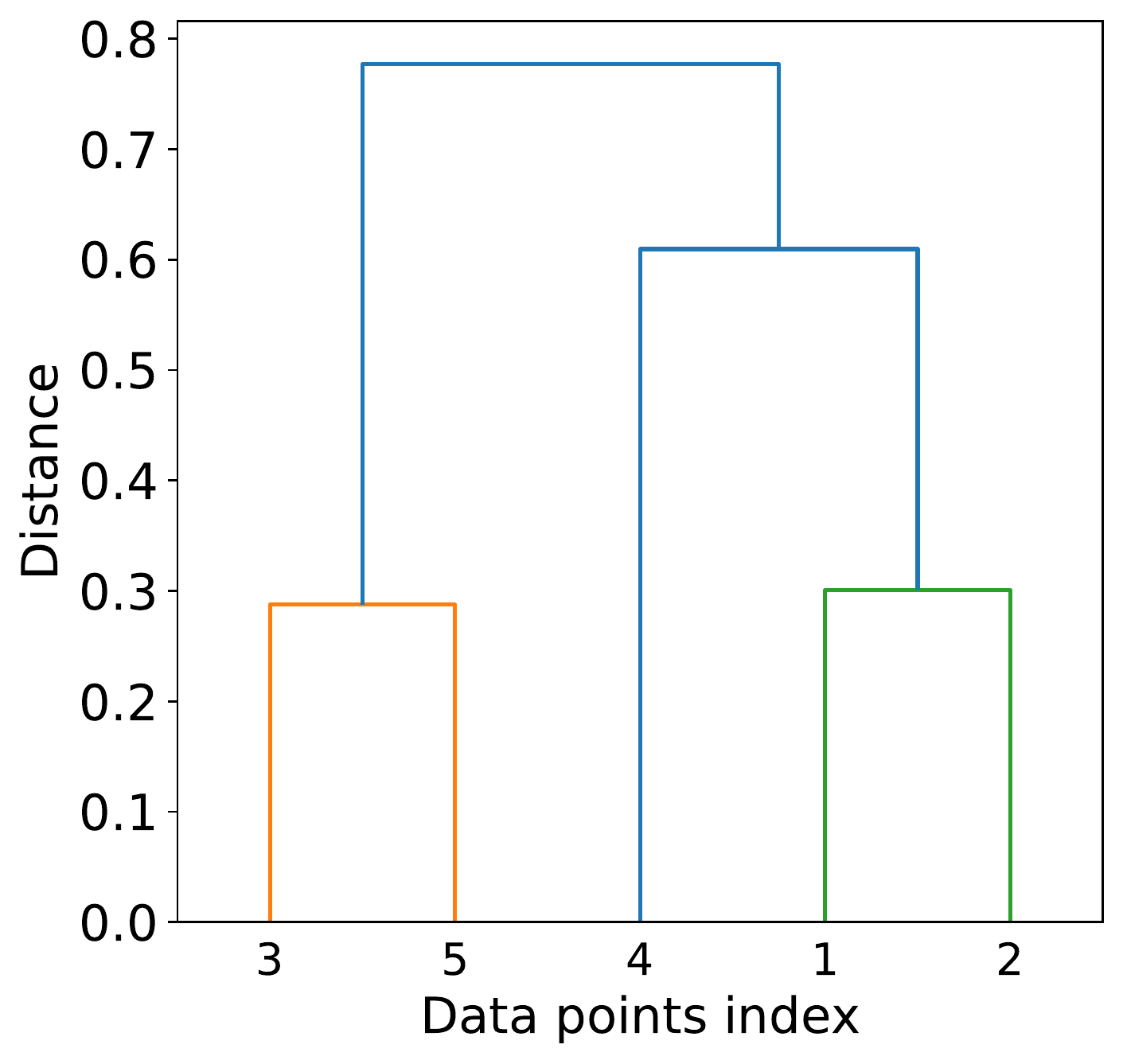}
    \caption{A sample dendrogram based on the result of the hierarchical clustering.}
    \label{fig:sample_dendrogram}
\end{figure}

After performing the hierarchical clustering, the dendrogram of the clustering results can be derived. Figure~\ref{fig:sample_dendrogram} shows an example of the dendrograms from clustering five nodes. As for a HES, after a disruptive event, the decision maker can determine the number of subsystems to form based on the specific system status. And the clustering results can be obtained by cutting the dendrogram at the corresponding level. For instance, if three clusters are preferred for the clustering results shown in Fig.~\ref{fig:sample_dendrogram}, then the islanding results can be obtained by cutting the tree at a level around $d=0.6$ to have three clusters $\{3,5\}$, $\{4\}$, and $\{1,2\}$. The advantage of obtaining the tree-like results for clustering is that it reveals a finer structure of the system, compared to directly outputting the partitions by using the traditional K-Means algorithm. At various levels of the clustering resolution, the practitioner can zoom in or out to examine different structures of the clusters, and follow how each cluster is formed hierarchically.

\subsection{Grassmann Manifold}\label{sec:grassmann}
Though the hierarchical clustering can expose a more detailed islanding results for HESs after disruptions comparing to traditional K-Means, the aforementioned algorithm has not considered the multiple Laplacian matrix defined simultaneously for the graph. In \cite{Sanchez-Garcia2014}, the authors demonstrate different intentional islanding results by using Laplacian matrices derived from different information, for instance, the average power flow and the admittance. But the scale of heterogeneous information embedded in the Laplacian matrix may be different, and it is required to normalize the magnitude to a unified range before executing the clustering process. Thus, how to combine all different kinds of electrical information of the power system remains to be answered. Rather than treating different types of the Laplacian matrices separately as in \cite{Sanchez-Garcia2014}, here we adopt the Grassmann manifold for multiple matrices in order to derive a unified Laplacian matrix for clustering.

A Grassmann manifold $G(K,N)$ can be defined as the set of $K$ dimensional linear subspaces in $\mathbb{R}^N$, where each distinct subspace is mapped to an unique point on the manifold. Therefore, information embedded in different spaces can be mapped into an unified space and the clustering algorithm can utilize the mapped information to derive islanding results. We summarize the key steps for finding a unified Laplacian matrix, or the Grassmann manifold of a graph with $M$ distinct Laplacian matrices as follows \cite{Dong2013}:
\begin{enumerate}
    \item Derive Laplacian matrix $L_i$ for each graph $G_i$ defined based on different weight matrices;
    \item Each $G_i$ can be approximated by the spectral embeddings matrix $U_i\in \mathbb{R}^{N\times K}$ from the first $K$ eigenvectors of $L_i$;
    \item Each $U_i$ can be considered as a $K$ dimensional subspaces in $\mathbb{R}^{N}$;
    \item The main objective becomes combing these several subspaces by finding a typical subspace $span(U)$, which is closest to all the subspaces $span(U_i)$; this can be written as a minimization program:
    \begin{equation}
        \min_{U\in \mathbb{R}^{N\times K}} \sum_{i=1}^M tr(U^TL_iU)+\alpha \left[kM-\sum_{i=1}^M tr(UU^TU_iU_i^T)\right];
    \end{equation}
    \item The solution to the above minimization problem is the unified Laplacian matrix for clustering:
    \begin{equation}\label{eqn:L_uni}
        L_{uni} = \sum_{i=1}^M L_i - \alpha \sum_{i=1}^M U_i U_i^T.
    \end{equation}
\end{enumerate}

\subsection{Choose the Embedding Space}
Once we have a unified Laplacian matrix derived from multiple heterogeneous Laplacian matrices as shown in Eqn.~\ref{eqn:L_uni}, then the intentional islanding can be achieved by performing hierarchical clustering on the Laplacian matrix as the global spectral embedding of the graph.

Although we don't need to specify the number of clusters $K$ beforehand to perform the hierarchical clustering, there is still a decision for choosing the subspace dimension $K$ to embed the network system. Without explicit information about the generator coherency, the number of desired dimensions $K$ is unknown and can be problem-dependent. But one practical approach to make the optimal decision for the subspace dimension is to calculate the eigenvalues of the Laplacian matrix and treat the eigengap as a metric. The eigengap is the difference between the consecutive eigenvalues and can be found by
\begin{equation}
    \gamma(\lambda_i) = |\lambda_{i+1}-\lambda_i|,
\end{equation}
where $\lambda_i$ is the i-th eigenvalue of the Laplacian matrix. And the normalized eigengap measured in percentage can be obtained from
\begin{equation}\label{eqn:eigengap}
    \gamma_n(\lambda_i) = \frac{\gamma(\lambda_i)}{\lambda_{i+1}}. 
\end{equation}
After deriving the eigengaps, how to determine the optimal $K$ remains to be discussed. Here we need to introduce the basic metric used to measure the quality of the intentional islanding without the electrical properties. Let's denote the volume of a cluster as $vol(A)=\sum_{i\in A, j\in A} W_{ij}$, where $A$ is a subset of the vertices that form the cluster. Based on the expression of the $Vol(A)$, one can see that it actually measures the closeness of the nodes inside the same cluster. Moreover, the boundary of the cluster is denoted as $b(A)=\sum_{i\in A, j\notin A}W_{ij}$, which measures the strength of interaction between the cluster and neighbors outside the cluster $A$. And the conductance $\Phi(A)$ of the cluster $A$ can be defined as the quotient $\frac{b(A)}{vol(A)}$. For an effective clustering strategy, the goal is to partition the graph into clusters with the smallest conductance. In other words, the optimal islanding results should only include clusters that have strong interactions within the cluster and weak connections to the rest of the graph outside the cluster. Formally, this clustering task can be written as the $k$-way partitioning problem \cite{Peng2017}. Multiple subsets of vertices, i.e., clusters, $A_1, \dots, A_k$ are a $k$-way partition of the original graph $\mathcal{G}$ if $A_i \cap A_j = \emptyset$ for different $i,j$ and $\cup_{i=1}^k A_i=\mathcal{V}$. The objective of the $k$-way partitioning problem can be defined by
\begin{equation}\label{eqn:expansion_constant}
    \rho(k):=\min_{A_1, \dots, A_k}\max_{1\leq i \leq k} \Phi(A_i).
\end{equation}
Following the definition of the objective, the partition tries to enforce that the conductance of any cluster $A_i$ is at most the $\rho(k)$, i.e., the $k$-way expansion constant. Finding the optimal partition that optimizes the $\rho(k)$ is generally not computationally efficient for large-scale networks. Moreover, solving the $k$ optimal cuts on a graph such as to search for the optimal solution of Eqn.~\ref{eqn:expansion_constant} is usually NP-hard \cite{Garey1976}. Nonetheless, the spectral clustering algorithm provides an approximation of the true optimum of the NP-hard problem. To be specific, Lee et al.~\cite{Lee2012} prove that performing spectral clustering on a graph $G$ provides a bound for the $k$-way expansion constant, defined by higher-order Cheeger inequality:
\begin{equation}\label{eqn:cheeger}
    \frac{\lambda_k}{2}\leq \rho(k) \leq O(k^2)\sqrt{\lambda_k},
\end{equation}
where $0\leq\lambda_1\leq\dots\leq\lambda_n\leq2$ are the sorted eigenvalues of the normalized Laplacian matrix of the original graph and $\lambda_1$ is the smallest eigenvalue. Equation~\ref{eqn:cheeger} suggests that a graph $\mathcal{G}$ can have clusters with good quality if and only if the eigenvalue $\lambda_k$ is small, which leads to a small $k$-way expansion constant. Whereas the practical approach for choosing the optimal $K$ is not only determined by the Cheeger inequality, which always gives the best result when choosing $K$ to be one. In general, $K$ is chosen to be the number associated with a large eigengap, and this approach has been proved by Lee et al.~\cite{Lee2012}. In their work they show that if there is a significant gap between the eigenvalues of $\mathcal{G}$, then Eqn.~\ref{eqn:cheeger} can be rewritten to be independent on $K$:
\begin{equation}\label{eqn:independent_cheeger}
    \max_{1\leq i \leq k} \Phi(A_i)\leq \sqrt{\lambda_k/\delta^3},
\end{equation}
where $\delta\in(0,\frac{1}{3})$, and $\lambda_k$ is the $k$-th smallest eigenvalue of the graph Laplacian $L$. That is, after obtaining the eigengap results from Eqn.~\ref{eqn:eigengap}, we need to choose the $K$ corresponding to a relatively large eigengap $\gamma_n(\lambda_k)$. This choice of $K$ not only leads to a small graph expansion constant with a small eigenvalue, but also provides a reasonable number of dimensions for the spectral embedding.

\subsection{Working with Renewables}\label{sec:renewables}
It's noteworthy to mention that all the Laplacian matrices defined in the previous sections are static and do not take the system's dynamic response into consideration. Moreover, the different behavior of the renewable generation is not covered. Thus in this section, how to derive the dynamic response of the renewable generation coupled in the power grid is discussed. In \cite{Khalil2018}, the authors have analyzed the major factor for deriving the dynamic response of the renewable generation. Take a wind power plant (WPP) as an example: if the WPP capacity is large, then the point of common coupling (PCC) transient response will be dominated by converter dynamics. In this case, the WPP will operate as a separate electrical area because of the distinct behavior different from nearby synchronous generators. However, if the WPP capacity is relatively low, then the WPP transient response at PCC is imposed by the rest of the power system and the WPP dynamic response is coherent with other generators. As a result, the PCC dynamics can be used as a proxy to identify the coherence of the renewable with other components inside the grid. Thus, this study adopts a measurement-based method and utilizes frequency deviation signals from system nominal frequency to identify generator coherency with the presence of renewable generations. And the coherency is treated as another type of Laplacian matrix for the generator dynamic information in the clustering algorithm.

The frequency deviation signal at each bus $i$ can be obtained by
\begin{equation}
    \Delta f_{i, t+\Delta t} = \frac{1}{\omega_0}\left(\frac{\phi_{i, t+\Delta t}-\phi_{i,t}}{\Delta t}\right),
\end{equation}
where $\omega_0$ is the system nominal frequency, and $\phi_i$ is the voltage phase angle at bus $i$. With the temporal frequency deviation signals, the frequency deviation vector of bus $i$ can be constructed from frequency deviation measurement at different time instants:
\begin{equation}
    \Delta f_i = \left[\dots,\Delta f_{i, t-\Delta t},\Delta f_{i, t},\Delta f_{i, t+\Delta t},\dots \right]^T.
\end{equation}
Once the frequency deviation vector of each bus after the disruption is obtained, the coherency between each pair of buses can be measured by the coherency coefficient $CC_{ij}$, which is derived from the Cosine similarity of the corresponding frequency deviation vectors:
\begin{equation}
    CC_{ij} = \frac{\Delta f_i \Delta f_j^T}{||\Delta f_i||||\Delta f_j||}.
\end{equation}
Since we can derive the $CC_{ij}$ for each pair of buses, or nodes in the graph, another weight matrix $W_{ij}^f$ carrying dynamic information for the power system graph can be defined:
\begin{align}
    W^f_{ij} = 
    \begin{cases}
    CC_{ij} &\text{if}\; (i,j)\in{\mathcal{E}} \\
    0 &\text{otherwise}
    \end{cases}
\end{align}

With all the above analysis, the overall road-map for conducting intentional islanding for HESs with renewable penetrations can be summarized as:
\begin{enumerate}
    \item Instead of finding the generator coherency first, use different information to define multi-layer graph Laplacian matrices for clustering
    \begin{itemize}
        \item Topology $w_{ij}=1\;\forall ij\in\mathcal{E}$, measures the pure connectivity
        \item Admittance $w_{ij}=1/\sqrt{R_{ij}^2+X_{ij}^2}\;\forall ij\in\mathcal{E}$, measures the strength of connections (electrical distance) 
        \item Average power flow $w_{ij}=(|P_{ij}|+|P_{ji}|)/2\;\forall ij\in\mathcal{E}$, measures the importance of a line during operation
        \item Frequency deviation $w_{ij} = CC_{ij}\;\forall ij\in\mathcal{E}$, measures the coherency of the buses, after a disruptive event
    \end{itemize}
    \item Construct the normalized Laplacian matrices and perform eigendecomposition. Analyze the eigengaps to determine the best number of clusters $K$, for deriving the Grassmann manifold
    \item Use Grassmann manifold to derive a unified Laplacian based on the multi-layer information:
    \begin{equation*}
        L_{uni} = \sum_{i=1}^M L_i - \alpha \sum_{i=1}^M U_i U_i^T
    \end{equation*}
    \item Conduct hierarchical clustering with connectivity constraints on the unified Laplacian
\end{enumerate}

\section{Results of Case Studies}\label{sec:case_study}
To validate the proposed clustering-based intentional islanding strategy, the IEEE 9-bus and IEEE 118-bus test systems are used for simulations. We have modified the original test networks to include additional renewable generations, such as wind farms, to show the applicability of the intentional islanding for transmission networks with renewables. The simulation of the system dynamic response after disruptions within a $20$ seconds time window is conducted in the Power System Simulator for Engineering (PSS/E) software. And line outages are simulated to happen at $2$ seconds. An idle time of $0.5$ second is imposed right after the occurrence of disruptions to simulate required response time of the system in real life.

\subsection{IEEE 9-Bus Test Network}
\begin{figure}
    \centering
    \includegraphics[width = 0.8\linewidth]{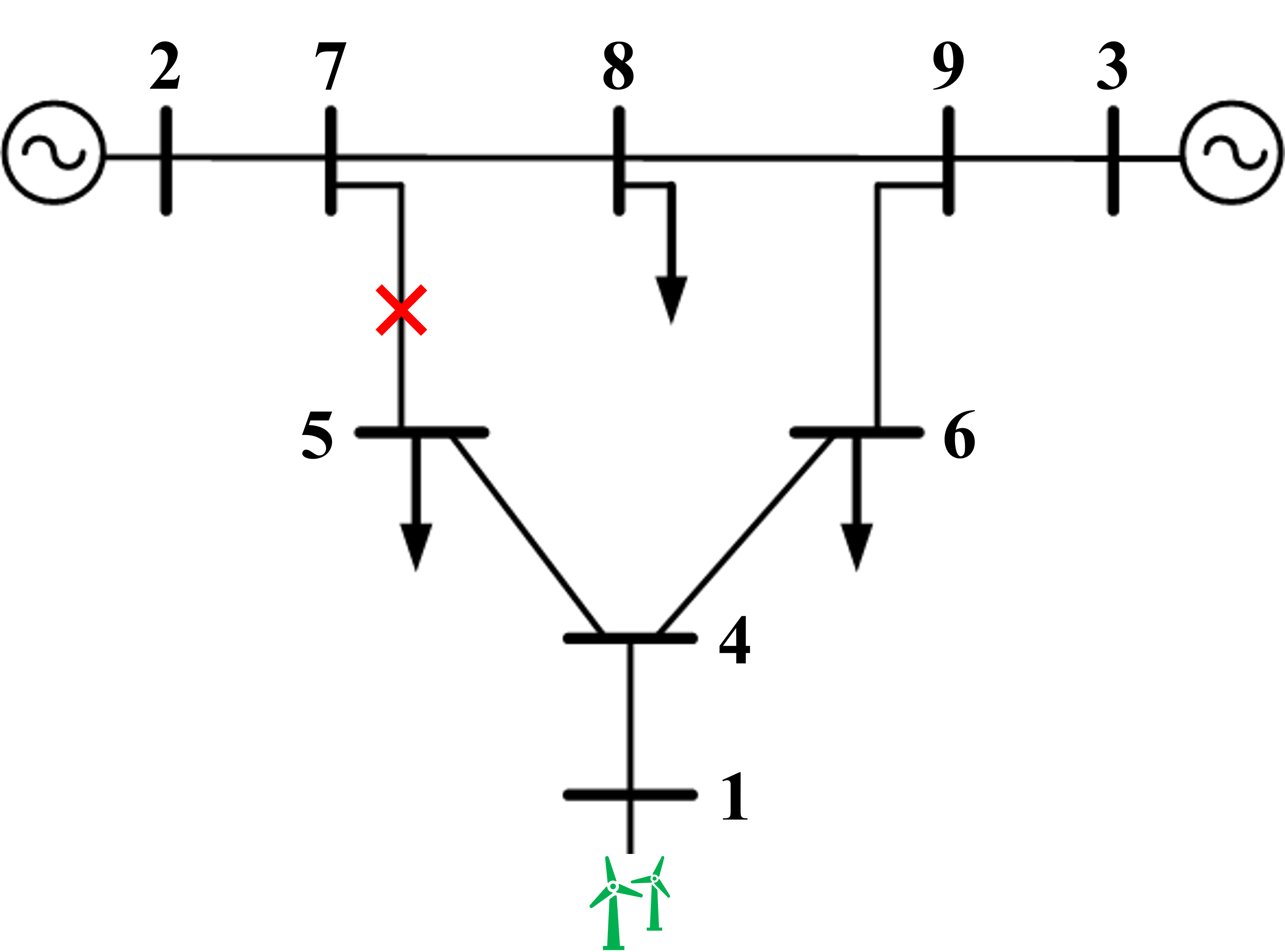}
    \caption{The one-line diagram of the IEEE-9 bus system, where the synchronous generator at bus 1 has been replaced by a wind farm (figure modified from \cite{song2015small}).}
    \label{fig:IEEE_9_bus}
\end{figure}
\begin{figure}
    \centering
    \includegraphics[width = 0.95\linewidth]{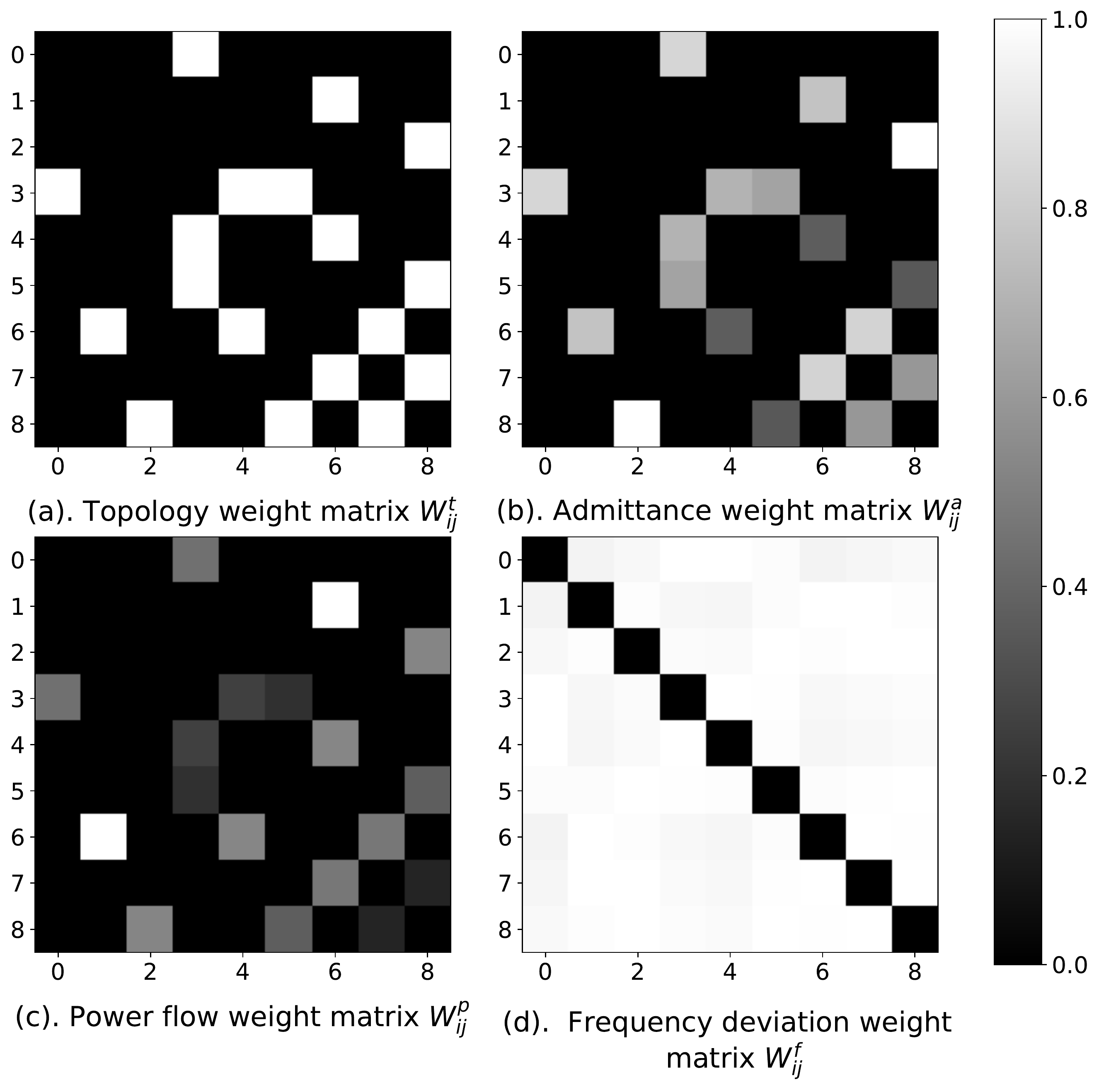}
    \caption{Four weight matrices constructed from the measurement data of the IEEE 9-bus system.}
    \label{fig:weight_matrices_9_bus}
\end{figure}
Firstly, to demonstrate the major ideas and steps of the proposed intentional islanding strategy, we use the IEEE-9 bus system as the baseline model to illustrate the details of the main steps of the algorithm. Figure~\ref{fig:IEEE_9_bus} shows the original layout of the IEEE 9-bus test case. The line outage happens at the transmission line $(7,5)$, in which the network has disconnected afterward. To incorporate the renewable generation, the synchronous generator at bus $1$ has been replaced by a wind farm. The capacity of the wind power generation is set to be $100$ MVA and consists of $30$ wind turbines. And this renewable generation corresponds to $33.3\%$ of the total power generation in the system. The original IEEE-9 bus system has a special characteristic that all three synchronous generators in the system have different operating modes, even before adding the renewable generation. Thus, this observation indicates that these three generators connecting at bus $1$, $2$, and $3$ need to be partitioned into separate islands with an effective intentional islanding strategy after disruptions. And because of this characteristic, studies about intentional islanding algorithms usually use the 9-bus system as the baseline model.

\begin{figure}
    \centering
    \includegraphics[width = 0.95\linewidth]{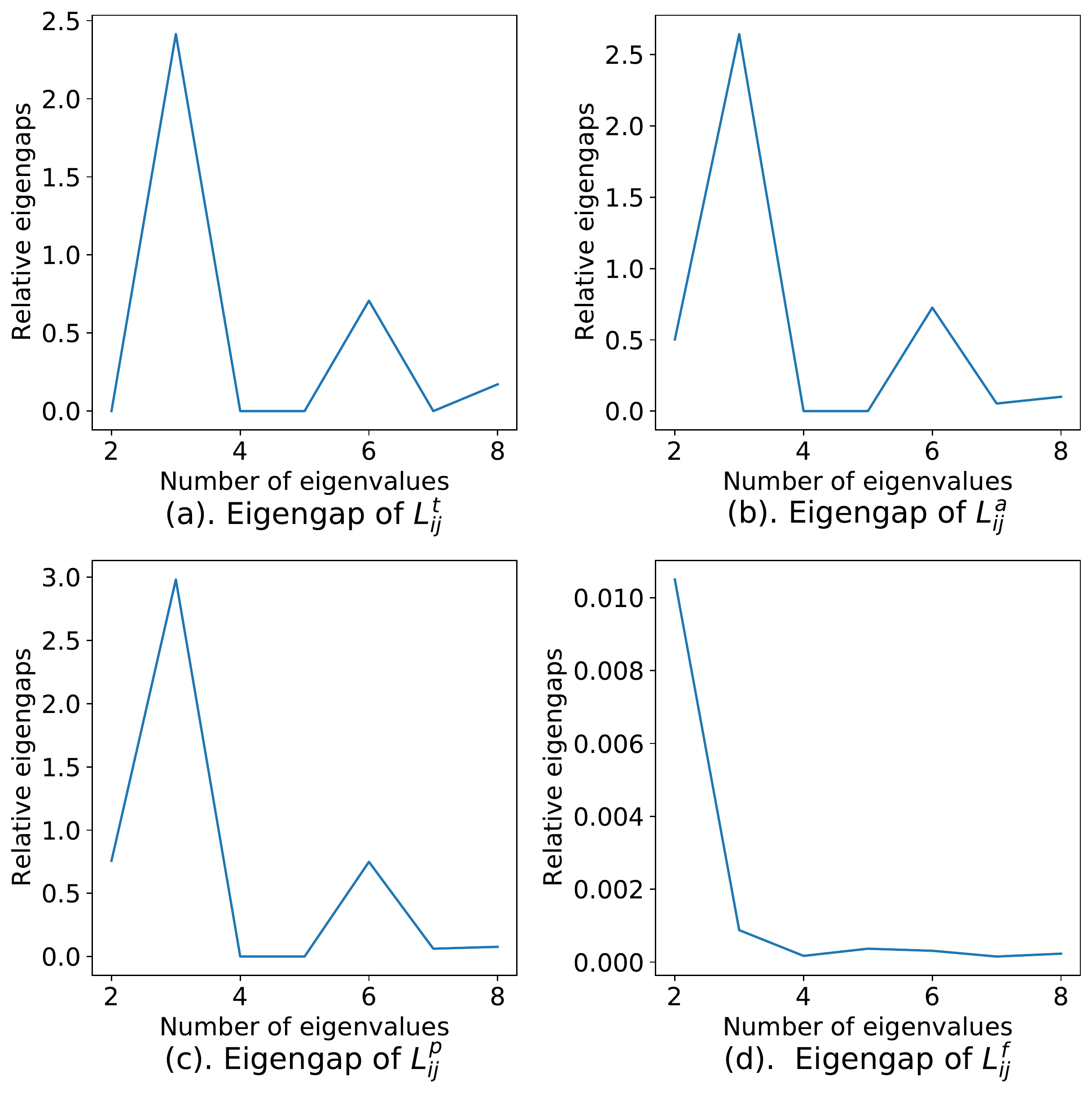}
    \caption{The eigengap obtained from the four Laplacian matrices defined based on topology, admittance, power flow, and frequency deviation.}
    \label{fig:eigengap_9_bus}
\end{figure}
\begin{figure}
    \centering
    \includegraphics[width = \linewidth]{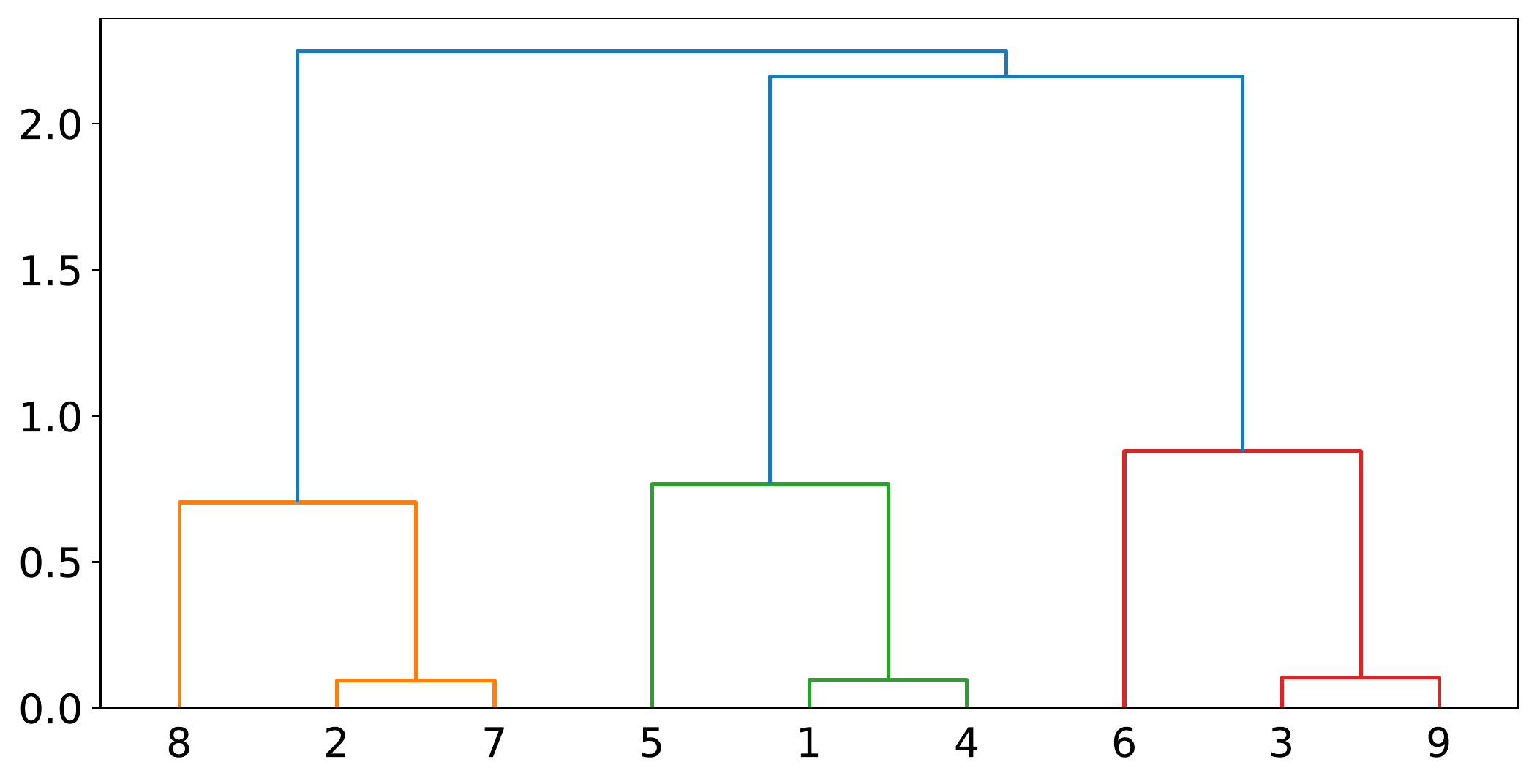}
    \caption{The dendrogram result of the hierarchical clustering for IEEE 9-bus system.}
    \label{fig:dend_9_bus}
\end{figure}
\begin{figure}
    \centering
    \includegraphics[width = 0.8\linewidth]{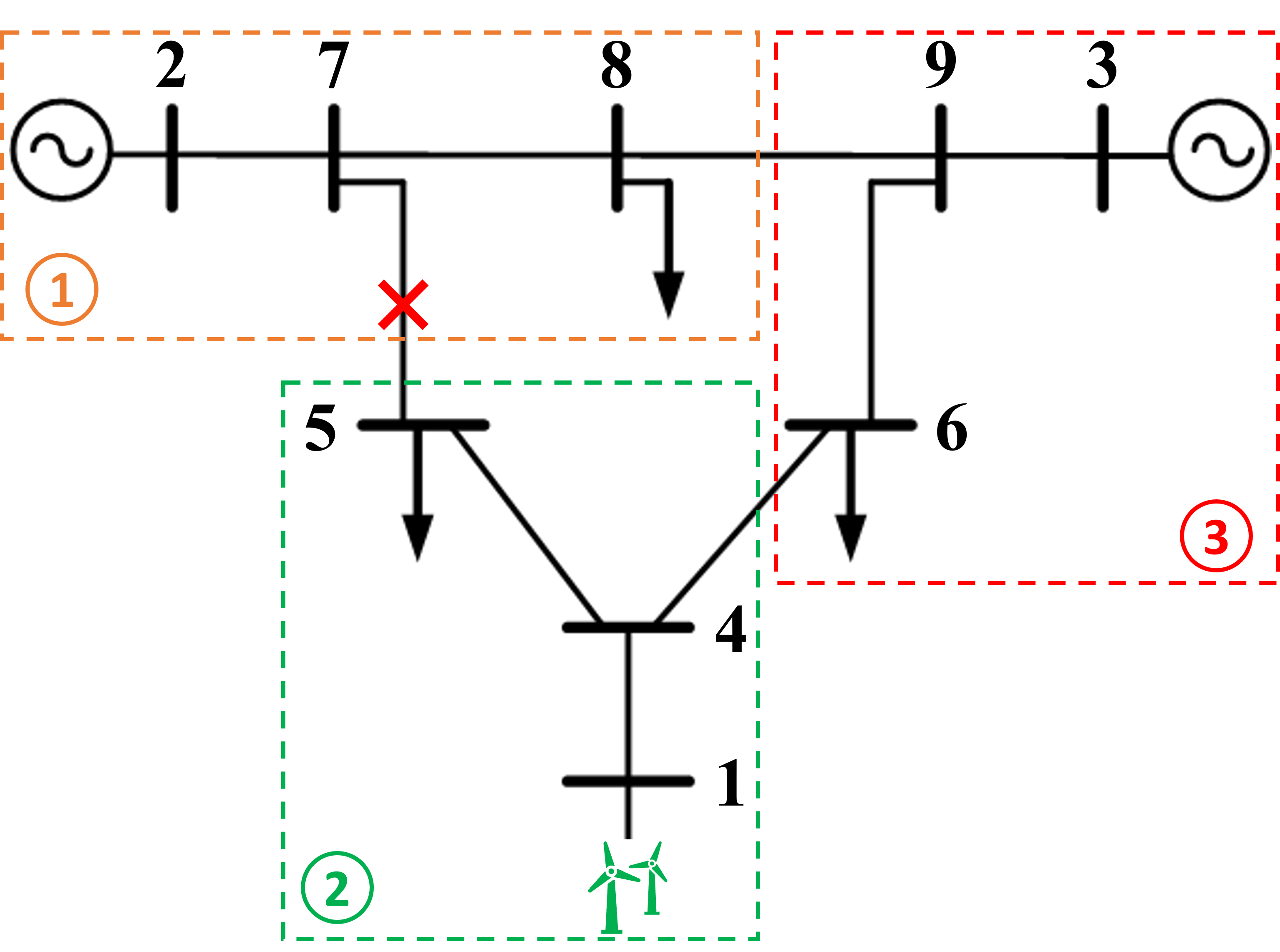}
    \caption{The intentional islanding results for the IEEE 9-bus system based on the derived dendrogram (figure modified from \cite{song2015small}).}
    \label{fig:islands_9_bus}
\end{figure}

As discussed in Sec.~\ref{sec:renewables}, the first step for the clustering-based intentional islanding is to define the weight matrices for the system from different types of measurements. For the IEEE 9-bus system, four weight matrices are constructed as shown in Fig.~\ref{fig:weight_matrices_9_bus}. From the matrices' data, it can be seen that except for the frequency deviation matrix, all other three matrices are sparse and share the same structure. This is due to the nature of connectivity: the power flow and the admittance can only be non-zero if there is a line between two buses. And a power system is indeed a sparse graph, where the number of edges has the same scale as the number of nodes, i.e., $|\mathcal{E}|\propto|\mathcal{V}|$. In contrast, the frequency deviation matrix is obtained from the measurement of the dynamic system response. And the deviation is a pairwise relation between buses regardless of the connectivity of the system. Figure~\ref{fig:weight_matrices_9_bus} shows that the frequency deviation matrix $W_{ij}^f$ has much more non-zero entries comparing to the topology, admittance, and power flow matrices. Thus, applying the Grassmann manifold to obtain a unified spectral embedding is a crucial task to conduct clustering because of the heterogeneous weight matrices.

Yet to derive the Grassmann manifold of the Laplacian matrices based on the weight matrices introduced above, we first need to determine the number of dimensions $K$ of the spectral embedding. As discussed in Sec.~\ref{sec:grassmann}, $K$ is determined to be the number of $i$, which gives a large eigengap between two consecutive eigenvalues of the Laplacian matrix $L$. Thus Fig.~\ref{fig:eigengap_9_bus} shows the eigengap results of the Laplacian matrix of the IEEE 9-bus system. Notice that the X-axis starts from the second eigenvalue since the smallest eigenvalue of any valid Laplacian matrix is always zero. Hence, from the plot, $K$ is chosen to be three in order to obtain the $K$ dimensional spectral embedding in the subspace for deriving the unified Laplacian matrix.

Figure~\ref{fig:dend_9_bus} demonstrates the clustering result of the IEEE 9-bus system. After a disruption, the system operator can obtain an appropriate intentional islanding strategy based on the dendrogram. For example, if three islands are desired, then three clusters can be formed by cutting the tree at around $1.0$ along the y-axis in the dendrogram. And the corresponding three clusters are the sub-trees that consist of buses $\{6,5,1,4\}$, $\{3,9\}$, and $\{8,2,7\}$. Figure~\ref{fig:islands_9_bus} illustrates the IEEE 9-bus system after forming isolated islands following the results in the dendrogram.
\begin{figure}
    \centering
    \includegraphics[width = \linewidth]{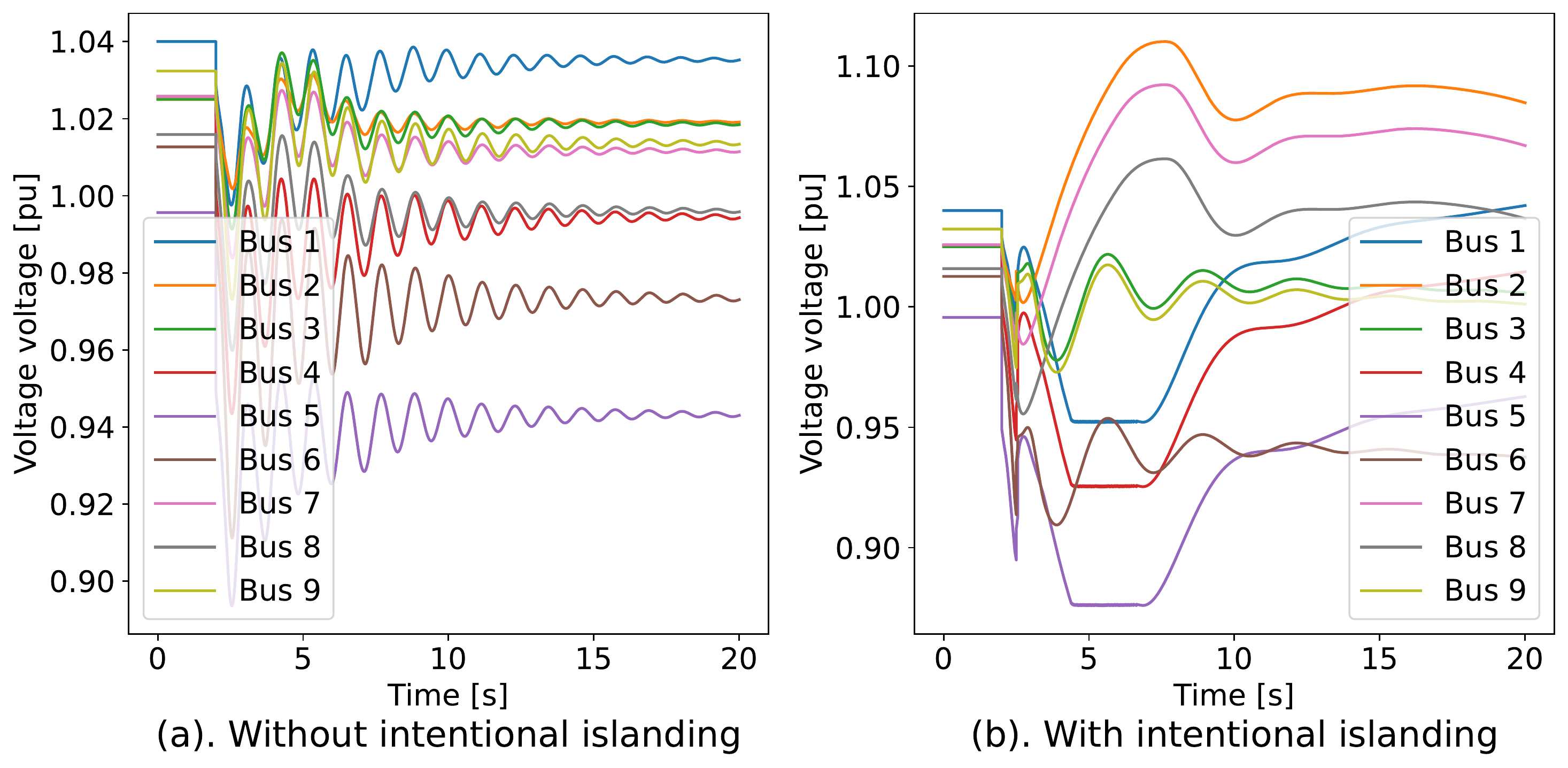}
    \caption{The dynamic response of the voltage magnitude at all buses of the IEEE 9-bus system after disruptions, with and without the intentional islanding operation.}
    \label{fig:voltage_response_9_bus}
\end{figure}
\begin{figure}
    \centering
    \includegraphics[width = \linewidth]{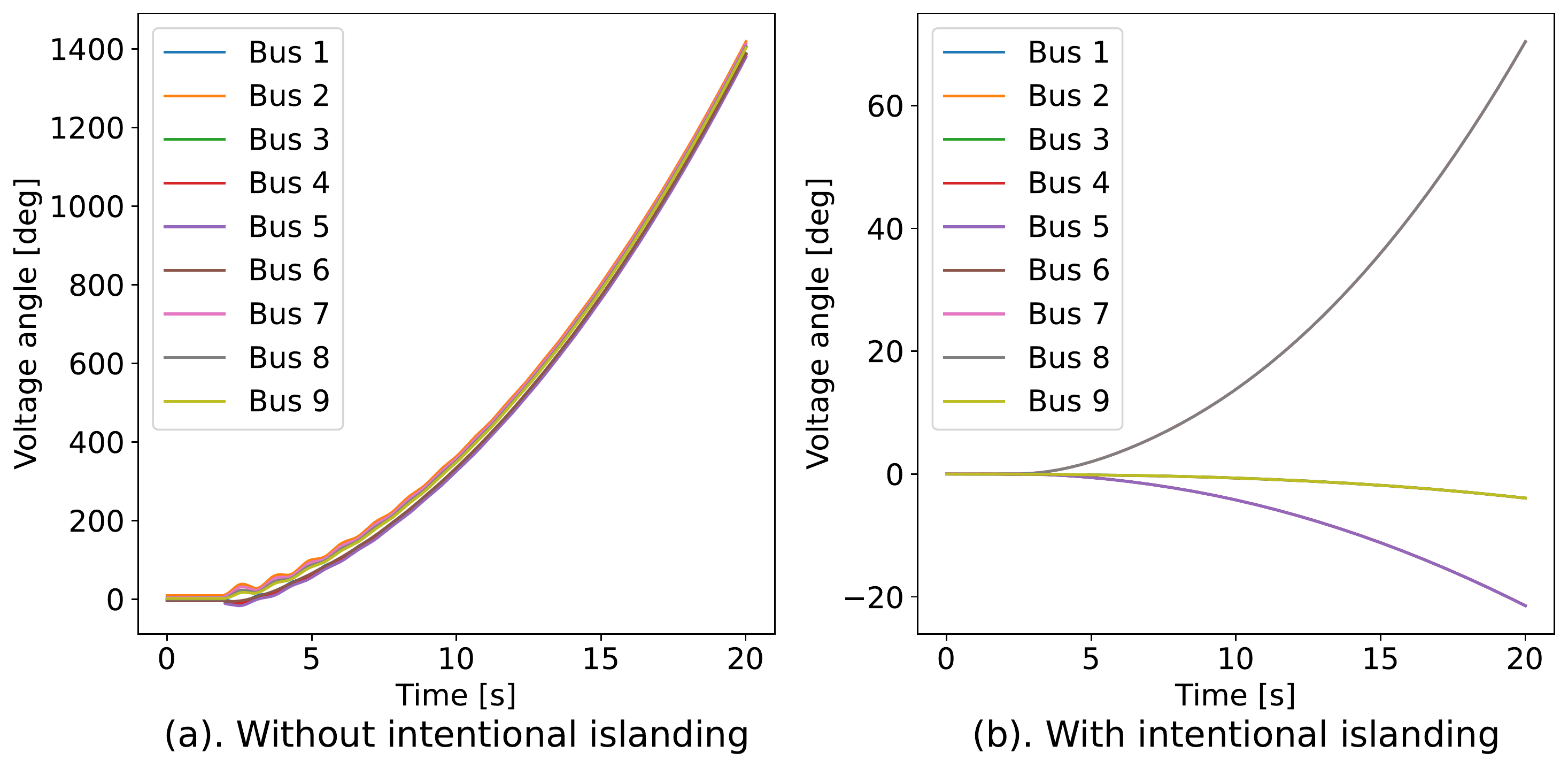}
    \caption{The dynamic response of the voltage angle at all buses of the IEEE 9-bus system after disruptions, with and without the intentional islanding operation: notice that the scale of the y-axis of the two plots are different.}
    \label{fig:ang_response_9_bus}
\end{figure}
To analyze the system dynamic response after the disruption with and without intentional islanding strategy, Fig.~\ref{fig:voltage_response_9_bus} and Fig.~\ref{fig:ang_response_9_bus} summarize the transient voltage magnitude and voltage angle measurements for the IEEE-9 bus test network, respectively. When the system undergoes a line outage without intentional islanding, Fig.~\ref{fig:voltage_response_9_bus} shows that the voltage magnitudes of all buses deteriorate from the nominal state to unstable cyclical patterns swinging between $0.92$ and $1.04$ pu. Although the voltage magnitudes have been self-stabilized toward the end of the simulation, the oscillations are so frequent that cascading failures maybe inevitable in practical applications. In comparison, with the intentional islanding strategy, the voltage measurements of all buses swing around $0.9$ to $1.1$ pu with much smaller frequencies and have been stabilized at new steady states around $10$ seconds. This difference is due to the fact that multiple lines, including the damaged line, are actively disconnected to enforce the system forming isolated sub-systems. And each sub-system is powered by one generator solely to ensure that the operating conditions are consistent across the buses inside the same island. 

Moreover, more significant difference can be found by examining the dynamic voltage angle response. Fig.~\ref{fig:ang_response_9_bus} suggests that the voltage angle of each bus becomes completely unstable while the angle can go beyond $1000$ degrees if no intentional islanding is conducted. On the contrast, with the help of the clustering results, the voltage angles stay within normal operating conditions, and the system can maintain nominal performance even with the presence of disruptions. This drastic difference in voltage angle response reveals the benefit of having the clustering-based intentional islanding as a contingency plan toward disruptions.

\subsection{IEEE 118-Bus Test Network}
\begin{figure}
    \centering
    \includegraphics[width = 0.8\linewidth]{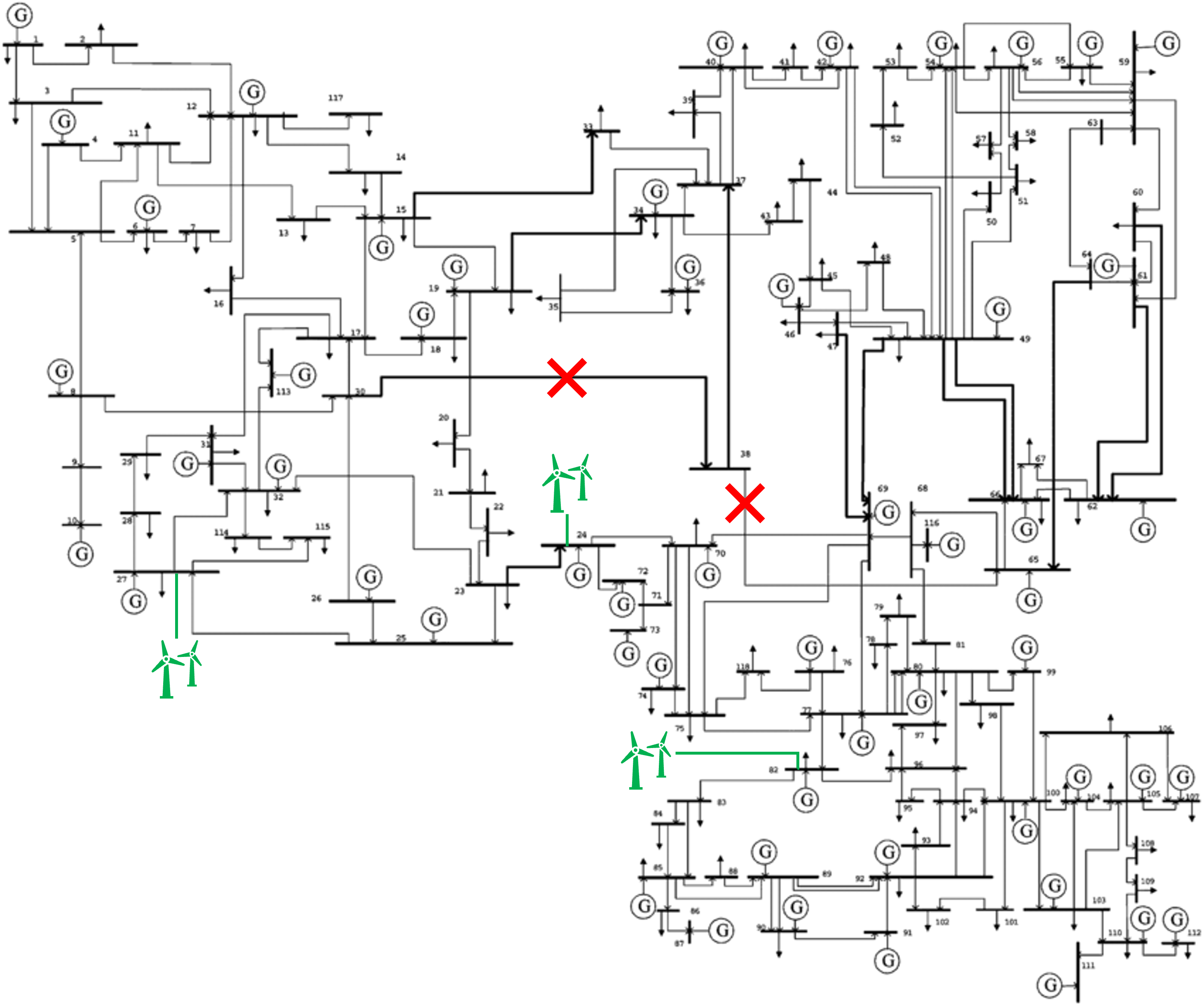}
    \caption{The one-line diagram of the IEEE 118-bus system, where three wind farms have been connected to buses 24, 27, and 82; line outages are simulated between buses 30/38, and between buses 38/65 (figure modified from~\cite{fernandez2018intentional}).}
    \label{fig:IEEE_118_bus}
\end{figure}
\begin{figure*}
    \centering
    \includegraphics[width = \textwidth]{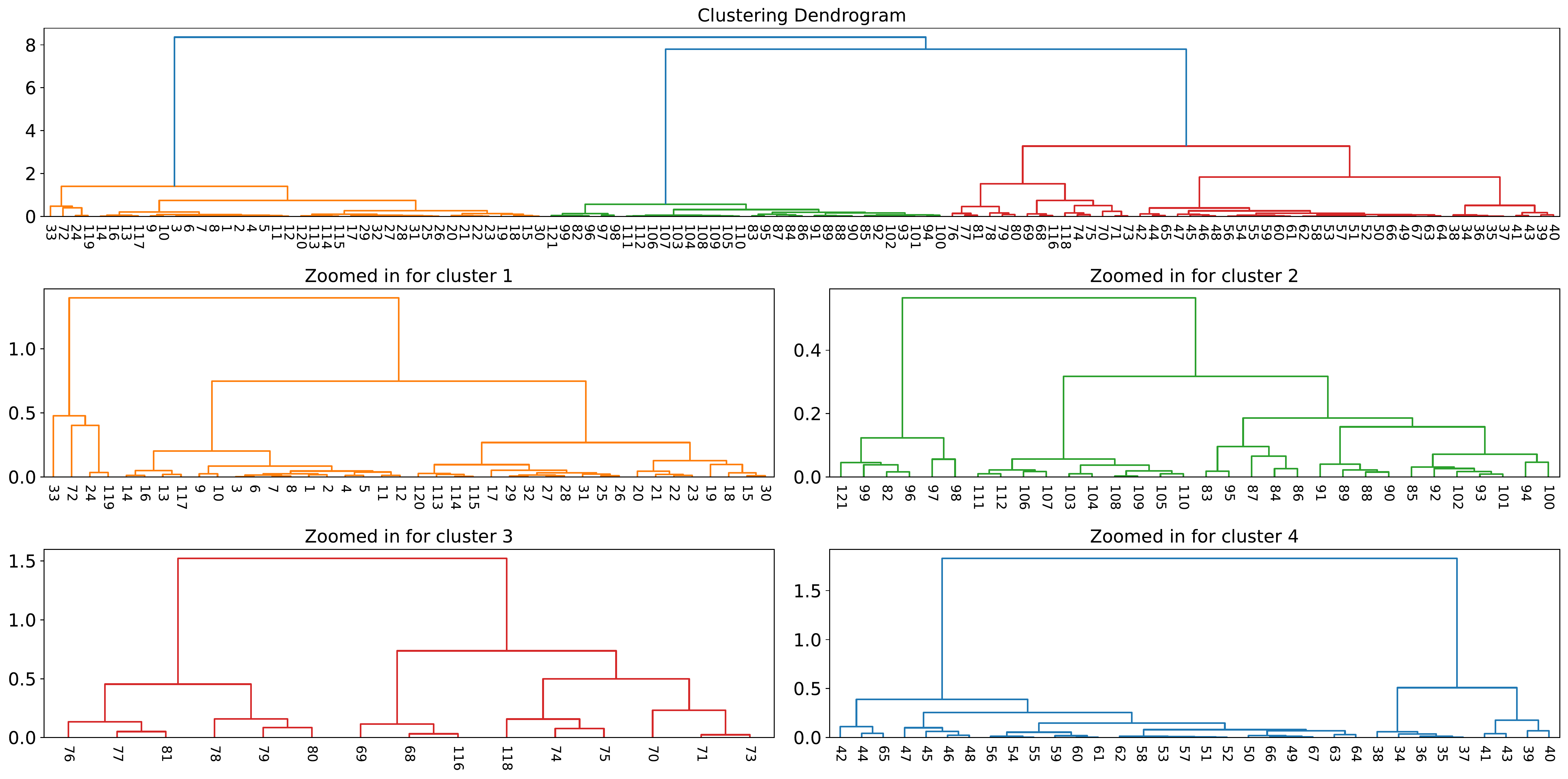}
    \caption{The dendrogram result of the hierarchical clustering for the IEEE 118-bus system: 4 clusters are chosen to be formed and their corresponding sub-dendrograms are included.}
    \label{fig:dend_118_bus}
\end{figure*}
\begin{figure}
    \centering
    \includegraphics[width = 0.8\linewidth]{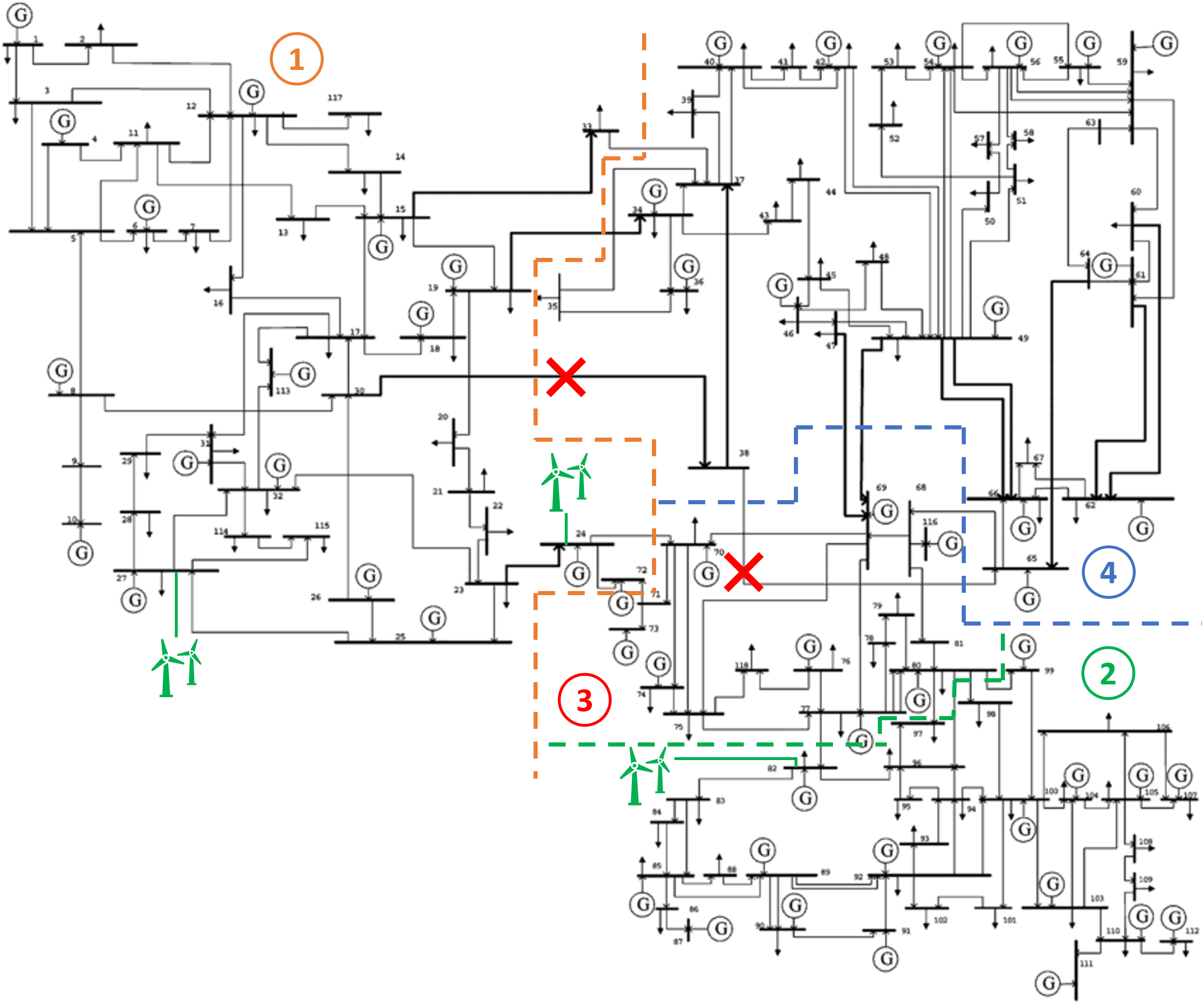}
    \caption{The intentional islanding results for the IEEE 118-bus system based on the derived dendrogram, where 4 clusters are formed (figure modified from~\cite{fernandez2018intentional}).}
    \label{fig:islands_118_bus}
\end{figure}

In addition to the small-scale IEEE 9-bus system, we demonstrate the effectiveness of the proposed intentional islanding algorithm with a large-scale IEEE 118-bus test network. To incorporate with the renewable generation, we have added three wind farms to the system at buses $24$, $27$, and $82$. All three wind farms have a generation capacity of 140 MVA. And these extra renewable generation corresponds to around 9.595\% of the total supply in the grid. The disruptive events are simulated to be line outages happening at $2$ seconds between the buses 30/38 and between the buses 38/65. The modified 118-bus system is illustrated in Fig.~\ref{fig:IEEE_118_bus}. After constructing the weight matrices and setting the dimension of the embedding subspace $K$ to be $3$, the dendrogram of the hierarchical clustering results is obtained and shown in Fig.~\ref{fig:dend_118_bus}. The clustering results shown in the dendrogram indicate that the modified IEEE 118-bus system forms $4$ subsystems after the disruption, depicted in four different colors. And the corresponding four isolated subsystems constructed from clustering results are illustrated in Fig.~\ref{fig:islands_118_bus}.
\begin{figure}
    \centering
    \includegraphics[width = \linewidth]{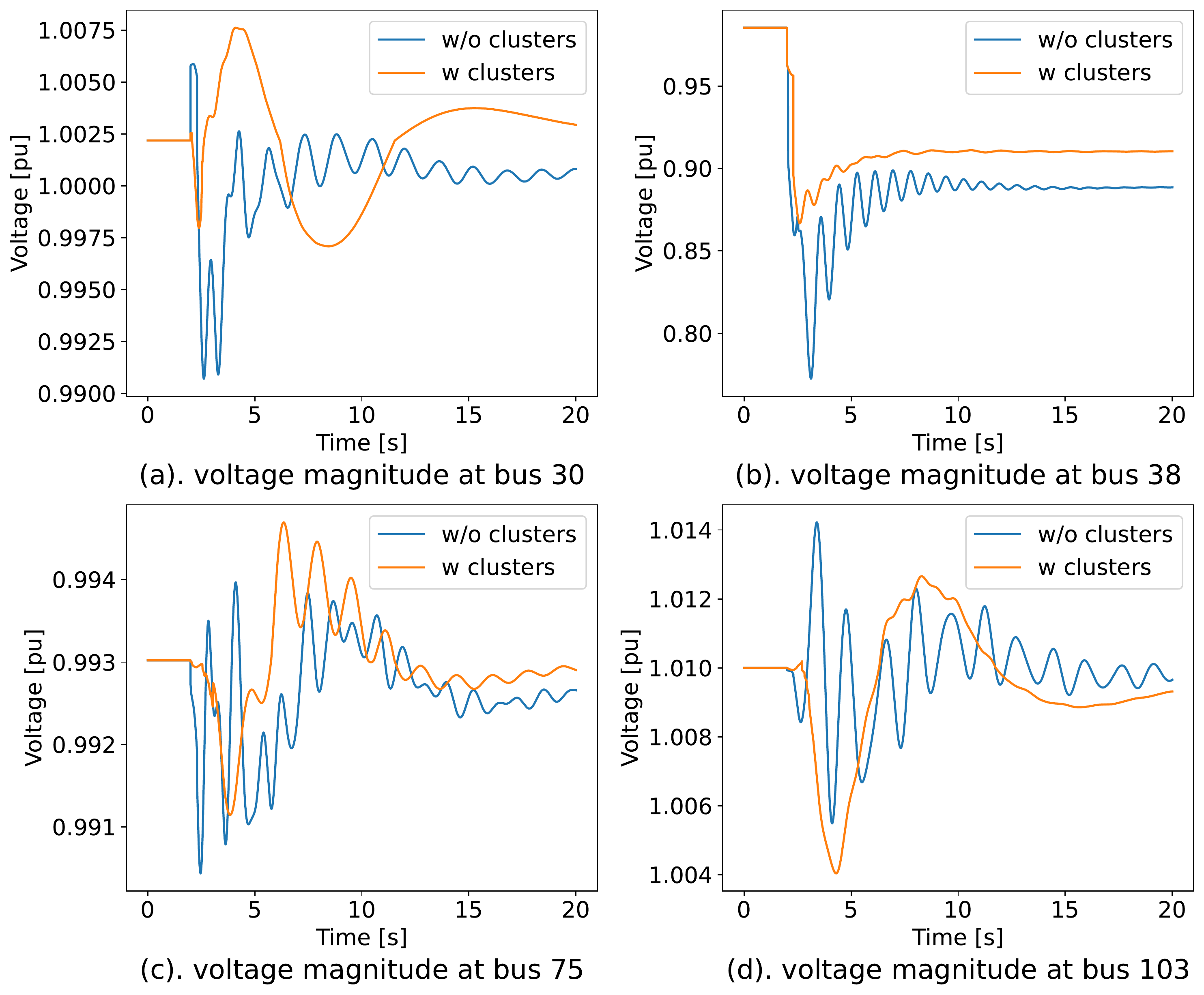}
    \caption{The dynamic response of the voltage magnitude at selected buses of the IEEE 118-bus system after disruptions, with and without the intentional islanding operation.}
    \label{fig:voltage_response_118_bus}
\end{figure}
\begin{figure}
    \centering
    \includegraphics[width = \linewidth]{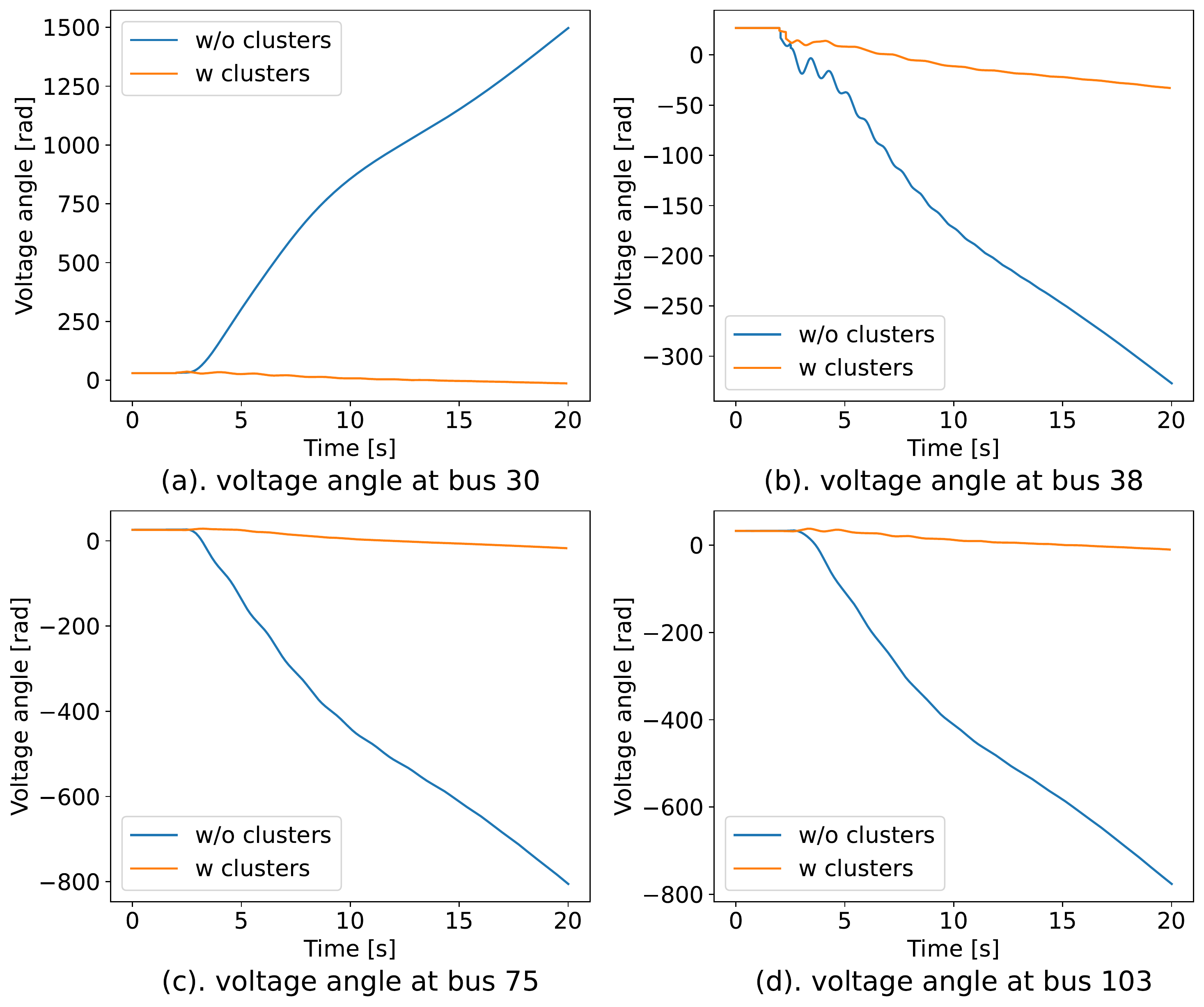}
    \caption{The dynamic response of the voltage angle at selected buses of the IEEE 118-bus system after disruptions, with and without the intentional islanding operation.}
    \label{fig:angle_response_118_bus}
\end{figure}

As for the system performance after disruptions with and without the intentional islanding control, we randomly pick four buses ($30$, $38$, $75$, and $103$) to study the system performance. And results of their dynamic voltage magnitudes and angles are shown in Fig.~\ref{fig:voltage_response_118_bus} and Fig.~\ref{fig:angle_response_118_bus}, respectively. After the line outage happens at $2$ seconds, the voltage magnitudes of all four buses begin to oscillate throughout the simulation. And this observation is similar to the small-scale IEEE 9-bus test case. After adding the intentional islanding strategy, the oscillations of the voltage magnitude indicate a damped phenomenon, though the amplitude changes insignificantly, especially for bus $30$, $75$, and $103$. This may due to the fact that nodes in the large scale HES, such as the 118-bus system, could be far away from the localized disruption and avoid severe downturn with the presence of external disturbance. Although the voltage magnitude results indicate that the large-scale 118-bus system can be stabilized automatically in terms of the voltages magnitude after the occurrence of the disruptive event, results of the voltage angle show the crucial importance for an intentional islanding strategy. In Fig.~\ref{fig:angle_response_118_bus}, the voltage angles of all four selected buses become out of synchronization and deviate indefinitely after injecting the line outage at $2$ seconds during the simulation. This significant deterioration in the voltage angle will eventually lead to cascading failure across the entire grid. In contrast, when the system is partitioned into different clusters according to the solutions of the clustering-based intentional islanding strategy, both the voltage magnitude and the voltage angle can be stabilized at appropriate level close to the pre-disruptive state. With the presence of the line outage, this stabilized performance demonstrates the benefit of adopting the intentional islanding process as a contingency plan to prevent the power system from cascading failures.

\section{Conclusion}\label{sec:conclusion}
To address the challenges of transmission system reliability along with the increasing level of renewable penetration, this study proposes a hierarchical clustering-based intentional islanding framework to guide the system transient operations after disruptions. Multiple layers of system real-time measurement data are considered including topology, electrical distance, as well as frequency deviation. To benefit from the heterogeneous information representing different aspects, this study adopts the Grassmann manifold algorithm to derive a unified embedding matrix for the final clustering. Moreover, because of the flexibility of the hierarchical clustering, the islanding decisions, i.e., the number of clusters formed, can be adjusted based on the stakeholders' prior knowledge. Also, to validate the proposed framework, the transient responses of the IEEE 9-bus and IEEE 118-bus systems after disruptions have been studied. However, in this study, we have only considered case studies with limited types of renewable generation, e.g., the wind farms with a relatively low level of penetration. In future studies, test systems with different types of generations are planned to be included, for instance, nuclear power plants with large generation capacities.

\appendices


\section*{Acknowledgment}
This research is partially supported by the National Science Foundation through the Engineering Research Center for Power Optimization of Electro-Thermal Systems (POETS) with cooperative agreement EEC-1449548, and the U.S. Department of Energy's Office of Nuclear Energy under Award No. DE-NE0008899.

\ifCLASSOPTIONcaptionsoff
  \newpage
\fi


\balance


\bibliographystyle{IEEEtran}
\bibliography{mybib}
\end{document}